\documentstyle[12pt]{article}
\makeatletter
\newcommand{\bd}{
\begin{document}}
\newcommand{\ed}{\end{document}}
\newcommand{\be}{\begin{eqnarray}}
\newcommand{\ee}{\end{eqnarray}}
\newcommand{\nn}{\nonumber}
\bd
\setlength{\baselineskip}{3ex}
\pagestyle{plain}
\pagenumbering{arabic}
\setcounter{page}{1}

\begin{center}
{\bf Some Phenomenology of the Top Quark with Non-standard Couplings}
$^{\ddagger}$\\
\end{center}
\begin{center}
Bing-Lin Young\\
Department of Physics \& Astronomy, Iowa State University\\
Ames, Iowa 50011$^\dagger$\\
Institute of Physics, Academia Sinica, Taipei, Taiwan\\
Institute of Theoretical Physics and Institute of High Energy Physics,
The Chinese Academy of Sciences, Beijing
\end{center}
\begin{abstract}
In this talk I will discuss possible new physics associated with the top quark.
We use higher dimension effective operators to represent the new physics and
examine constraints on the operators from the phenomenology they predicted.\\
$\ddagger$ Based on a talk presented at the International Symposium on Heavy
Flavor and Electroweak Theory, 16-19 August, 1995, Beijing, China.\\
$\dagger$ Permanent address
\end{abstract}

\noindent
{\bf I. Introduction}

Because of the large mass of the top quark which signifies a sizable coupling
to the electroweak symmetry breaking sector, it is possible that the top quark
may play a key role in probing new physics.  The works described below, which
are performed in collaboration with Seungkoog Lee, Kerry
Whisnant, and Xinmin Zhang, include the investigation of higher dimension
operators and flavor changing neutral currents, both can be remnants of the
physics took placed above the fermi scale. We examined constraints on the
effective operators due to baryogenesis, collider physics, unitarity bounds,
and certain static properties of leptons and hadrons. Due to space limitation,
only a bare number of references will be given.

I will first summarize the constraints on physics at the fermi scale due to
electroweak baryogenesis which is related to physics above the fermi scale.
The observed ratio of baryon number density to entropy,
             $\frac {n_B}{s} = (0.4 - 1.4)\times 10^{-10}$,
can impose significant constraints on the standard model. Two constraints are
particularly relevant to our discussion and both require the existence of new
physics. One is that there should exist new sources of CP violation besides
the complex phase of the CKM quark mixing matrix. With the CKM phase, the
electroweak dynamics predicts a very small baryon number density which is of
10 orders of magnitude smaller than the observation. A second constraints is
the bound on the Higgs boson mass.  In order to prevent washout of the baryon
number generated by the electroweak dynamics during the electroweak phase
transition, the Higgs boson mass can not exceed $45 GeV$ \cite{shap1}.
However, this bound conflicts with the LEP data \cite{particledata} which put
$58 GeV$ as the lower limit of the Higgs boson mass. \\
\\
{\bf II. Effective Lagrangian}

New physics by the effective Lagrangian is a familiar approach
\cite{effectLag}. Since the effective Lagrangian is free from specific models,
it is a systematic way in searching for new physics.  We assume that the
non-standard top quark coupling has something to do with symmetry breaking,
and is therefore associated with Higgs interactions.  We have investigated in
some detail the following lowest dimension operator,\\
$${\bf O} = \frac{m_t}{v}c_t{e^{i\xi}}\frac{{\mid \Phi \mid}^2
           - v^2/2}{\Lambda^2}
   ({\overline t}_l, {\overline b}_l){\tilde\Phi}t_r,$$\\
where $c_t$ and $\xi$ are unknown parameters.  Adding this operator to the
standard model Lagrangian, we obtain a modified top quark-Higgs boson
coupling,\\
$$\it L_H = {\frac{m_t}{v}}H
            {\overline t}\{1 + \delta(cos\xi + isin\xi)+ \cdots\}t, \, \,
  \delta = c_t\frac {v^2}{{\Lambda}^2}.$$\\
In the following, we summarize the effect of the operator on several
measurable quantities which allow us to constrain the parameters
$c_t$ and $\xi$ and make predictions.\\
\\
(II.A)  {\bf Extension of the Higgs mass bound} \cite{xmz1} \cite{xmzbly}

One effect of the operator is to increase the bound of the Higgs mass from
$M^2_{H0}$ to $M^2_{H0} + \frac{8 v^2}{\Lambda^2} v^2$, where $M^2_{H0}$
is subjected to the old bound of $45 GeV$. The new bound can be as large as
$180 GeV$ for $\Lambda = 1 TeV$.  This new bound may also imply that the
current LEP data allows a cutoff no higher than $\Lambda = 3 TeV$.\\
\\
(II.B)  {\bf The baryon number to entropy ratio} \cite{xmzbly}

The baryon number density to entropy ratio can be estimated as
$\frac{n_B}{s} \approx \kappa{\alpha}^4_W {\delta}_{CP}F,$
where $F$, being of the order of $0.1$, depends on the properties of the
electroweak phase transition, and $\kappa$ governing the rate of baryon
number violation in the symmetric phase ranges from 0.5 to 20. The CP
violation phase from the operator {\bf O} is given by $\delta_{CP} \sim
c_t{sin\xi}\frac{v^2} {2{\Lambda}^2}$.  Using $\Lambda = 1 TeV$, we have
$\kappa c_t sin\xi \geq 4 \times 10^{-2}.$

It should be noted that the electroweak baryogenesis calculation is only
accurate to within a couple of order of magnitude.  For instance, if the
effect of QCD sphaleron is taken into account, the predicted baryon asymmetry
will be suppressed by a factor of $10^{-2}$ and the above bound will be
$\kappa c_t sin\xi \geq 4.$ \\
\\
(II.C) {\bf Electric dipole moments of electron and neutron} \cite{xmzbly}

The operator ${\bf O}$ can contribute to the electric dipole moments of the
fermion.  First, a CP violating $H\gamma\gamma$ coupling can be generated
through the top quark loop diagram. Then the photon coupling to the fermion
under consideration through a virtual H and $\gamma$ one-loop process will
produce the desired electric dipole moments, \\
$$d_f \sim e\cdot Q_f^2\frac{m_f}{v^2}(\frac{\alpha}{27\pi})
   c_t{sin\xi}\frac{1}{16\pi^2}ln{\frac{m_t^2}{m_H^2}},$$
where $Q_f$ is the electric charge of the fermion in units of e.  Using the
constraints on the input parameters of the effective operator, we obtain
$d_e \sim \frac{6}{\kappa}(10^{-28} - 10^{-30})e{\cdot}cm$,
where the experimental data is $d_e^{exp} = (-0.3 \pm 0.8)\times
10^{-26}e{\cdot}cm$.  The neutron electric dipole moment is given by
$d_n \approx (\frac{m_d}{m_e}) d_e
  \sim \frac{1}{\kappa}(10^{-26}-10^{-28})e{\cdot}cm$,  and the
experimental upper limit is $d_n^{exp} < 11 \times 10^{-26} e{\cdot}cm$.
Both are one to two orders of magnitude below the experimental limit.\\
\\
(II.D)  {\bf Phenomenology at high energy linear colliders} \cite{xmzbly}

The effective operator {\bf O} has interesting consequences at the future
high energy $e^+e^-$ linear collider.  For example, the cross section of
$e^+ + e^- \rightarrow t + \overline t + H$ is sensitive to the parameters
of {\bf O}.  For the top quark mass of $180 GeV$ and Higgs boson mass
$100 GeV$, about 20 events per year are predicted by the SM to be produced for
an integrated luminosity of $\int \L = 20 fb^{-1}$. The event can be
identified by the spectacular final state
$W^+W^-b{\overline b}b{\overline b}$.  Except for special values of the
parameters the cross section value in the presence of the anomalous coupling
can be distinguished from that of the standard model.  We plotted the cross
section versus $\delta$ and $\xi$.  {\bf Fig. 1}.\\
\\
(II.E) {\bf Bounds from unitarity} \cite{kwbly}

We have also considered the unitarity constraints on the anomalous top quark
Yukawa coupling.  We performed a multichannel analysis of the helicity
amplitudes of the six reactions $t \overline t \rightarrow$
$t_{\pm} \overline t_{\pm}$, $W_L^+ W_L^-$, $Z_L Z_L$, $Z_L H$, and $HH$.
In the analysis we also took into consideration the $t{\overline t}HH$
anomalous vertex contained in {\bf O}.  We plot the constraints of
$\delta$ versus $\xi$ from the unitarity, baryogenesis, and the electric
dipole moment of the neutron. {\bf Fig. 2}.

Let us remark that an operator like {\bf O} can be realized in a left-right
model when the heavy right-handed degree of freedom is integrated out
\cite{xmzbly}. \\
\\
{\bf III. Flavor changing neutral current}

Non-universal electroweak interactions which couple only to the
third generation
can lead to flavor changing neutral currents (FCNC) due to quark mixing
\cite{rdp}.  The
strength of the FCNC depends on the individual quark mass mixing matrix
elements.  We can write the FCNC in the form
$$j_{\mu}^{jk} = \overline q_j\gamma_\mu(g^{eff}_{jk,L}\Gamma_L
                                       + g^{eff}_{jk,R}\Gamma_R)q_k$$
$$g^{eff}_{jk,C} = g_C(\delta_{jk} + \kappa_{jk,C}),$$
C = L (R) means left- (right-) handed, $\Gamma_L$ ($\Gamma_R)$ are the
left- (righ-) handed chiral projection operator, $\kappa_{jk,C}$ gives the
strength of the flavor changing neutral current, and
$g_L = -\frac{1}{2} + \frac{1}{3}sin^2\theta_W$,
$g_R = \frac{1}{3}sin^2\theta_W$ are the standard model couplings
of $Z$ to ${\overline b} b$.
The $\kappa$ term can be induced by higher dimension effective operators.
Since $g_R$ is much smaller than $g_L$, we will neglect the effect of the
right-handed part. We take the following operators \cite{xmzbly2}
$${\it O_1} = i[\Phi^\dagger{D_\mu}\Phi - (D_\mu\Phi)^\dagger \Phi]
            \overline\Psi_L\gamma^\mu\Psi_L$$
$${\it O_2} = i[\Phi^\dagger \vec{\tau} {D_\mu}\Phi
               - {(D_\mu\Phi)^\dagger} \vec{\tau} \Phi]
            \overline\Psi_L{\gamma^\mu} \vec{\tau} \Psi_L,$$
where $\overline\Psi_L = (\overline t_L, \overline b_L)$ is the left-handed
third generation doublet. The effective Lagrangian is given by
$$ L_{eff} = L_{SM} + \frac{1}{\Lambda^2}(c_1{\it O_1} + c_2{\it O_2})$$
After the spontaneous symmetry breaking and diagonalization of the quark
mass matrix, an anomalous neutral current is obtained,
 \[ \frac{g_2}{cos\theta_W} {\left( \begin{array}{ccc} \overline d, &
   \overline s, & \overline b \end{array} \right) }_L U_L^{(d) \dagger}
   \left( \begin{array}{cccc} 0 & & \\ & 0 & \\ & & \Delta_L
   \end{array} \right) U_L^{(d)} \gamma_\mu
   {\left( \begin{array}{c} d\\ s\\ b \end{array} \right) }_L Z^\mu \]
where $g_2$ is the $SU(2)$ coupling,
$\Delta_L = \frac{v^2}{\Lambda^2}(c_1 + c_2)$, and $U_L^{(d)}$ is the unitary
rotation matrix which diagonalizes the left-handed down quark mass matrix.
Note that the CKM mixing matrix is given by
$V_{CKM} = U_L^{(u) \dagger}U_L^{(d)}$, where $U_L^{(u)}$ is the up quark
rotation matrix. We see that the FCNC depends on both the new physics and the
quark rotation matrix.

Since the individual quark rotation matrix is not known.  We used a variant
\cite{dsdzzx} of the modified Fritzsch ansatz of the down quark rotation
matrix
\[ \left( \begin{array}{lll} 1 & -(\frac{m_d}{m_s})^{1/2} &
  (\frac {m_d m_s (m_s + w_q)}{m^3_b})^{1/2}\\
  (\frac{m_d}{m_s})^{1/2} e^{-i\alpha_q} & e^{-i\alpha_q} &
  (\frac {m_s + w_q}{m_b})^{1/2} e^{-i\alpha_q}\\
  -(\frac{m_d(m_s + w_q)}{m_s m_b})^{1/2} e^{-i(\alpha_q + \beta_q)} &
  -(\frac{m_s + w_q}{m_b})^{1/2} e^{-i(\alpha_q + \beta_q)} &
  e^{-i(\alpha_q + \beta_q)} \end{array} \right) \]
where $w_q$ is taken to be $m_c$, $\alpha_q$ and $\beta_q$ are responsible
for the CP violation phase in the CKM mixing matrix.

The above effective Lagrangian can affect a number of low energy reactions.
We refer to \cite{xmzbly2} for details.

We are investigating, in another approach to FCNC, the anomalous electro- and
chromo-magnetic operators, $tc\gamma$ and $tcg$. The effective
Lagrangian is taken to be
$${\Delta}L = \frac{1}{\Lambda}
             (\kappa^{\gamma}e{\overline t}{\sigma}_{\mu\nu}c F^{\mu\nu}
 + \kappa^g g_s{\overline t}
{\sigma}_{\mu\nu}\frac{{\lambda}^j}{2}cG^j_{\mu\nu})$$
Both $\kappa^{\gamma}$ and $\kappa^g$ can be constrained\cite{kwbly2}
from $b \rightarrow s + \gamma$ \cite{buras}  \cite{hewett}
which has been observed experimentlly
\cite{bsgamma}. The anomalous $tc\gamma$ coupling contributes to $bs\gamma$
directly through a one-loop diagram and $tcg$ contributes through QCD
corrections.

The $t \rightarrow c + \gamma$ decay is easily observable by the presence of
a high energy $\gamma$ which carries half of the top quark energy.  There
seems to be very little background.  We are performing a simulation to study
the background in detail.  The process considered is the $t \overline t$
production in a high energy hadron collider.  One of the top quarks decays
into $c + \gamma$ and the other decays through the normal channel
\cite{kwbly2}. A sizable anomalous $tcg$ couplng can enhance the single top
quark production at the Fermilab Tevatron which we are currently
investigating.

Part of this work was performed at the Institute of High Energy Physics,
Beijing, the Istitute of Theoretical Physics, Beijing, and the Insttitute of
Physics, Taipei during the author's sabbatical leave. The author would like
to thank Professors Tao Huang, Jimin Wu, Yuanben Dai, Zhaobin Su, Zhaoxi
Zhang, Zhongyuan Zhu, Hai-Yang Cheng, and Shih-Chang Lee for generous
supports and for the gracious hopitalities extended to him.

\ed